\documentclass[12pt,noshowpacs,nofootinbib,notitlepage,amsmath]{revtex4-1}
\allowdisplaybreaks
\usepackage{graphicx,color}
\usepackage[colorlinks=true,citecolor=blue,linkcolor=blue,urlcolor=blue]{hyperref}
\usepackage[charter]{mathdesign}
\DeclareSymbolFontAlphabet{\mathcal}{symbols}
\DeclareSymbolFont{symbols}{OMS}{xmdcmsy}{m}{n}
\DeclareSymbolFont{largesymbols}{OMX}{xmdcmex}{m}{n}
\SetSymbolFont{symbols}{bold}{OMS}{xmdcmsy}{b}{n}

\begin{document}  %\color[rgb]{0.3,0.3,0.5}
\title{\color{blue}\Large On Dark Matter Selected  High-Scale Supersymmetry}
\author{Sibo Zheng}
\email{sibozheng.zju@gmail.com}
\affiliation{Department of Physics, Chongqing University, Chongqing 401331, P.R. China}
\begin{abstract}
The prediction for the Higgs mass in the dark matter selected high-scale SUSY is explored. 
We show the bounds on SUSY-breaking scale in models of SM $+\tilde{w}$ and SM $+\tilde{h}/\tilde{s}$ 
due to the observed Higgs mass at the LHC. 
We propose that effective theory below scale $\tilde{m}$ described by SM $+\tilde{w}$  
is possibly realized in gauge mediation with multiple spurion fields that exhibit significant mass hierarchy,
and that by SM $+\tilde{h}/\tilde{s}$ can be realized with direct singlet-messenger-messenger coupling for singlet Yukawa coupling  $\lambda\sim(v/\tilde{m})^{1/2}g_{\text{SM}}$.
Finally, the constraint on high-scale SUSY is investigated in the light of inflation physics 
if these two subjects are directly related.
\end{abstract}
\maketitle

\section{Introduction}
In particle physics the interest on supersymmetry (SUSY) is based on four main reasons: 
$i)$ solution of the naturalness problem, 
$ii)$ successful gauge coupling unification, 
$iii)$ viable thermal dark matter (DM) candidate with mass near weak scale, 
$iv)$ ingredients of string theory.
The so called low-scale SUSY addresses $i)$-$iii)$.
However, the first run of LHC shows bad prospect for conventional low-scale SUSY.
In contrast to low-scale SUSY, only $iv)$ is addressed in the ``minimal '' high-scale SUSY,
it is because of that in ``minimal '' high-scale SUSY all superpartner masses are far above the weak scale,
and the connection between weak and SUSY breaking scale is lost.
But $iii)$ should be addressed in any realistic model,
and this can be done in two classes of  ``non-minimal '' high-scale SUSY 
\footnote{Actually, issue $ii)$ can be roughly addressed in no-minimal high-scale SUSY discussed here,
and it is possible to achieve unification of gauge couplings in high precision. }.

The first one is named as ``split'' SUSY \cite{splitsusy}, 
in which the scalar superpartners masses are far above the weak scale, 
but all fermionic superpartners including gaugino $\tilde{g}$ 
and higgsinos $\tilde{h}_{u}$ and $\tilde{h}_{d}$  are light due to the protection of $R$ symmetry.
In this class of high-scale SUSY DM  is identified as the lightest supersymmetric particle.
The observed Higgs mass in high precision \cite{higgsmass}, 
by virtue of two-loop RGEs and one-loop threshold corrections, 
suggests that the scale of SUSY breaking  $\tilde{m}\leq 10^{8}$ GeV \cite{1108.6077,1407.4081} 
when scalar superpartner threshold corrections are not very large.
Above the scale $\tilde{m}$ the physical states are described as the 
minimal supersymmetric standard model (MSSM),
while below it described as standard model (SM) $+\tilde{g}+\tilde{h}$.

The other class was firstly studied in \cite{0912.3942},
in which $R$-symmetry breaking isn't suppressed, 
and either some new parity instead of $R$ symmetry keeps higgsino ($\tilde{h}$) and singlino ($\tilde{s}$) light or there exists a light wino $\tilde{w}$ DM due to environmental selection.
In the former case, the singlino state is needed because of that pure higgsino DM isn't viable.
So below the scale $\tilde{m}$ the physical states are described as SM$+\tilde{h}/\tilde{s}$ (SM$+\tilde{w}$)
when DM is mixing state of $\tilde{h}$ and $\tilde{s}$ ($\tilde{w}$-like). 
The observed Higgs mass, 
similar to  the analysis performed in the first class,
can be used to constrain the scale of SUSY breaking  $\tilde{m}$.
Since the region of model parameters discussed in \cite{0912.3942}  
corresponds to Higgs mass of order  $127-142$ GeV in model SM$+\tilde{w}$ 
and  $141-210$ GeV in model SM$+\tilde{h}/\tilde{s}$ (see Table 4 therein) 
it is necessary and also interesting to revise the Higgs mass in such DM selected high-scale SUSY.
This is the aim of this paper.
In particular, instead of taking $\tilde{m}=10^{14}$ GeV and large $\tan\beta$ limit as in  \cite{0912.3942} , 
 $\tilde{m}$ will be considered as a free parameter in this paper, 
and region of small $\tan\beta$ will be covered also.

The paper is organized as follows.
Similar to Split SUSY \cite{splitsusy}
we discuss the two-loop RGEs for Higgs quartic coupling $\lambda $ 
and SM gauge and Yukawa couplings in subsection IIA.
The threshold correction to $\lambda$ due to heavy SUSY particles 
will be parameterized for the prediction for Higgs mass. 
In subsection IIB to IIC,  we estimate the prediction for Higgs mass $M_h$ in models 
SM$+\tilde{w}$ and SM$+\tilde{h}/\tilde{s}$, respectively.
The uncertainty for $M_h$ due to uncertainty of top quark mass and/or 
threshold correction will be emphasized.
Section III is devoted to the model building for high-scale SUSY studied in this paper.
In section IV we discuss the possible constraint on high-scale SUSY from the inflation physics in the early universe.
Finally, we conclude in section V.
Two-loop RGEs for parameters related to Higgs mass are presented in appendix A.

\section{Higgs Mass}
In this section, we estimate the prediction for the Higgs mass.
Similar to the case of split SUSY \cite{1407.4081},
we use the updated experimental values of top quark mass $M_{t}=173.3 \pm 0.76$ GeV \cite{topmass} 
and QCD coupling $\alpha_{3}(M_{Z}) = 0.1184 \pm 0.0007$ \cite{g3} for our analysis.
The measured value for the Higgs mass, $M_{h}=125.15\pm 0.25$ GeV 
is obtained from a naive average of the ATLAS and CMS results \cite{higgsmass}.
As the Higgs mass is rather sensitive to top Yukawa,
the dominant one-loop QCD corrections to top Yukawa $\delta y_{t}\simeq -0.065$ \cite{1108.6077, 1301.5167} will be applied to derive the prediction.

\subsection{RGEs and Threshold Corrections}
As the Higgs mass is directly related to $\lambda$, 
we will use the two-loop RGEs for all relevant couplings in SM$+\tilde{w}$ and SM$+\tilde{h}/\tilde{s}$.
Inspired by the study of split SUSY in \cite{1108.6077} and high-scale SUSY in \cite{1203.5106},  
the two-loop RGEs for our model parameters can be similarly derived,
which are presented in appendix A.
A few comments are in order.
$i)$, The two-loop beta functions for SM couplings 
can be derived by either following \cite{Machacek, Luo} or taking the insights of Weinberg \cite{weinberg}.
$ii)$, In the light of the model SM $+$ Majorana triplet fermion (T) $+$ Dirac doublet fermion studied in  \cite{1203.5106}, the correction to RGEs in SM$+\tilde{w}$ is derived by decoupling the Dirac doublet fermion and identifying $T$ as wino which is a triplet $(\chi^{\pm},\chi^{0})$.
$iii)$, A new parameter $g_{\lambda}$ appears in model SM$+\tilde{h}/\tilde{s}$ (see Eq.(\ref{lagrangian})).
In the light of the model SM $+$ Majorana singlet fermion $+$ Dirac doublet fermion (D) studied in \cite{1203.5106}, the correction to RGEs is derived by identifying Dirac doublet fermion $D$ and $D^{c}$ as higgsinos $\tilde{h}_{u,d}$.

\linespread{1}\begin{figure}[!h]
\centering
\begin{minipage}[b]{0.75\textwidth}
\centering
\includegraphics[width=4.5in]{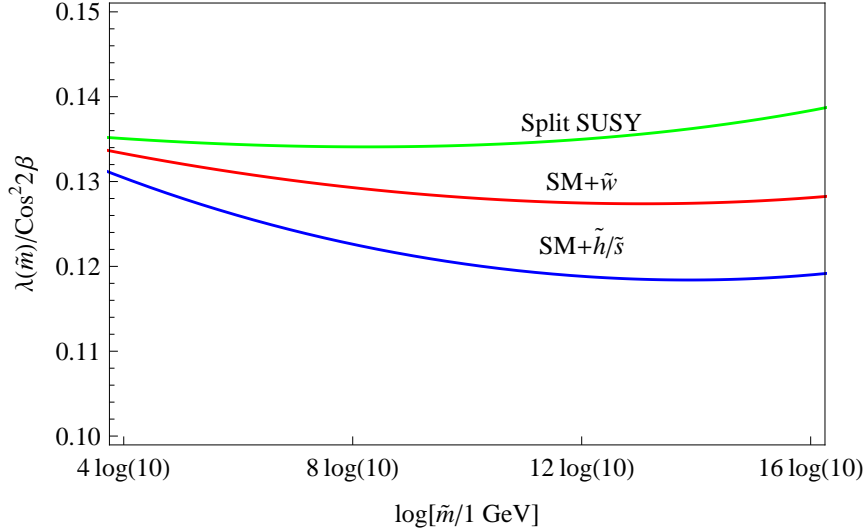}
\end{minipage}%
\caption{$\lambda(\tilde{m})/\cos^{2}2\beta$ as function of $\tilde{m}$ for $\delta\lambda^{(\text{SUSY})}=0$.}
\label{boundary}
\end{figure}

The value of Higgs quartic coupling at scale $\tilde{m}$,  $\lambda(\tilde{m})$, 
is determined by the SUSY boundary condition, 
\begin{eqnarray}{\label{higgsmass}}
\lambda(\tilde{m})=\frac{1}{4}\left[g_{2}^{2}(\tilde{m})+\frac{3}{5}g^{2}_{1}(\tilde{m})\right]\cos^{2} 2\beta
+\delta \lambda^{(\text{SUSY})}(\tilde{m}).
\end{eqnarray}
up to threshold correction $\delta\lambda^{(\text{SUSY})}$,
which arises from the heavy SUSY particles. 
Solving the RGE for $\lambda$ and taking the SM threshold correction $\delta\lambda^{(\text{SM})}$ into account,
one obtains the electroweak (EW) scale Higgs mass $M_{h}=2\lambda v^{2}$, 
with the EW scale $v=174$ GeV. 

The threshold correction arising from SM particles has been calculated in \cite{threshold}. 
See also \cite{1307.3536} for the full two-loop SM correction.
The numerical value for $\delta\lambda^{(\text{SM})}$ is given by,
\begin{eqnarray}{\label{deltasm}}
\delta\lambda^{(\text{SM})}\simeq -\frac{G_{F}M^{2}_{Z}}{8\sqrt{2}\pi^{2}}\left(\xi F_{1}+F_{0}+F_{3}/\xi\right)\lambda\simeq 0.0075 \lambda,
\end{eqnarray} 
where $G_{F}$ denotes the Fermi constant for muon decay, 
functions $F_{i}$ are defined in \cite{threshold}, 
and $\xi=M^{2}_{h}/M^{2}_{Z}$.
Threshold correction arising from SUSY particles has been considered 
at one-loop level in \cite{1407.4081,0910.2235} and at the two-loop level in \cite{0910.2235}.
In the next section, we consider the range $\mid \delta\lambda^{(\text{SUSY})}\mid\leq 0.03$,
which is sufficient to include the uncertainty of theoretic value of $\delta\lambda$ due to heavy SUSY particle contribution.

Fig.\ref{boundary} shows the relative value $\lambda$ defined in Eq.(\ref{higgsmass})
for split SUSY,  SM$+\tilde{w}$ and SM$+\tilde{h}/\tilde{s}$,
by using the one-loop RGEs for SM gauge couplings (see appendix A).
It indicates that in split SUSY the value of $\lambda$ 
and the prediction for Higgs mass at EW scale is roughly the largest among the three models 
for the same $\tan\beta$ and $\delta\lambda^{(\text{SUSY})}$.
An exception is that the correction due to RGE effects is large enough to violate the expectation above.

\subsection{SM$+\tilde{w}$}
In model SM$+\tilde{w}$ the model parameters are wino mass $m_{\tilde{w}}$ and $\tilde{m}$.
Mass parameter $m_{\tilde{w}}$ is constrained by the DM relic abundance,
from which $m_{\tilde{w}}\sim 2$ TeV  \cite{0912.3942}.
Parameter $\tilde{m}$ relates to the boundary value for the Higgs mass at high energy scale,
and thus the measured Higgs mass is sensitive to it. 
Fig. \ref{wino1} shows the prediction for Higgs mass as function of $\tilde{m}$ by using two-loop RGEs (shown in appendix A) for $\tan\beta=\{1, 2, 4, 50\}$,
with the solid lines correspond to the central values $M_{t}=173.3$ GeV and $\alpha_{3}(M_{Z})=0.1184$.
The dotted lines represent the uncertainty of the prediction due to uncertainty of top quark mass.
The horizontal band indicates the measured value for the Higgs mass.
With $\tan\beta$ fixed, the uncertainty of Higgs mass at high energy scale shrinks to be about $\sim 1-2$ GeV at EW scale.
When threshold corrections are small, $\delta\lambda^{(\text{SUSY})} \sim 0$,
the model can be allowed in the wide range $ 10^{4}$ GeV $\leq\tilde{m}\leq 10^{16}$ GeV.

\linespread{1}\begin{figure}[!h]
\centering
\begin{minipage}[b]{0.75\textwidth}
\centering
\includegraphics[width=4.5in]{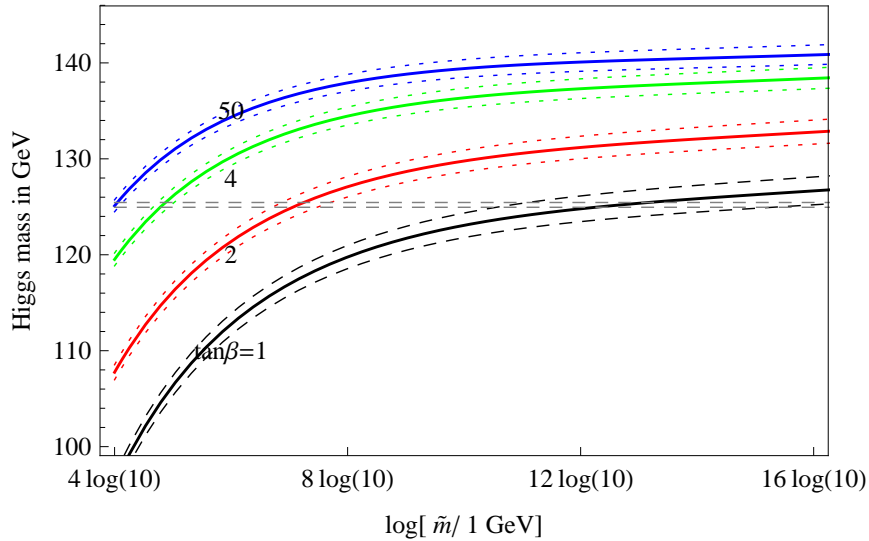}
\end{minipage}%
\caption{Higgs mass as function of SUSY-breaking scale $\tilde{m}$ in SM$+\tilde{w}$ by using two-loop RGEs, 
for $\tan\beta=\{1, 2, 4, 50\}$.
The solid curves corresponds to the central values $M_{t}=173.3$ GeV and $\alpha_{3}(M_{Z})=0.1184$,
while the dotted curves in each color show the uncertainty of the prediction due to experimental uncertainty in $M_{t}$.
Threshold corrections $\delta\lambda^{(\text{SUSY})}$ are assumed small and ignored.}
\label{wino1}
\end{figure}

\linespread{1}\begin{figure}[!h]
\centering
\begin{minipage}[b]{0.75\textwidth}
\centering
\includegraphics[width=4.5in]{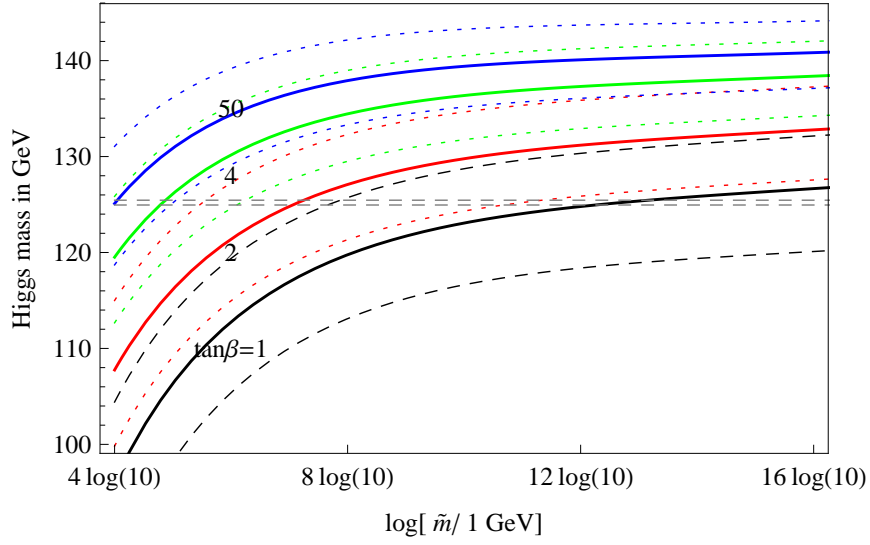}
\end{minipage}%
\caption{Same as Fig.\ref{wino1} but with threshold corrections $\mid\delta\lambda^{(\text{SUSY})}\mid\leq 0.03$ taken into account.}
\label{wino2}
\end{figure}

Compare our prediction for Higgs mass in Fig.\ref{wino1} with previous result in \cite{0912.3942, 0910.2235}.
$M_{h}$ approaches to $\sim 140$ GeV for large value of $\tan\beta$,
which is consistent with the prediction of $M_{h}\simeq 141-142$ GeV in \cite{0912.3942, 0910.2235}.
On the other hand, 
the prediction for Higgs mass should be similar to the minimal high-scale SUSY studied in \cite{1407.4081},
because the deviation from SM is smaller than split SUSY.

Fig.\ref{wino2} shows the case for threshold correction taken into account.
The solid line in each color corresponds to the central values of $M_{t}(M_{Z})$ and $\alpha_{3}(M_{Z})$ and $\delta\lambda^{(\text{SUSY})}=0$.
The dotted lines in each color represent the deviation from above due to the uncertainty of $M_{t}(M_{Z})$ and $\delta\lambda$.
This figure has shown that the uncertainty of $M_h$ is about $\sim 10$ GeV for $\mid \delta\lambda^{(\text{SUSY})}\mid \simeq 0.03$ in compared with Fig.\ref{wino1}.
It clearly shows that $\tilde{m}\leq 10^{8}$ GeV is allowed for $\delta\lambda^{(\text{SUSY})}=0.03$.
We expect that this bound increases for smaller threshold correction $\delta\lambda^{(\text{SUSY})} <0.03$.

\subsection{SM$+\tilde{h}/\tilde{s}$}

\linespread{1}\begin{figure}[!h]
\centering
\begin{minipage}[b]{0.5\textwidth}
\centering
\includegraphics[width=3in]{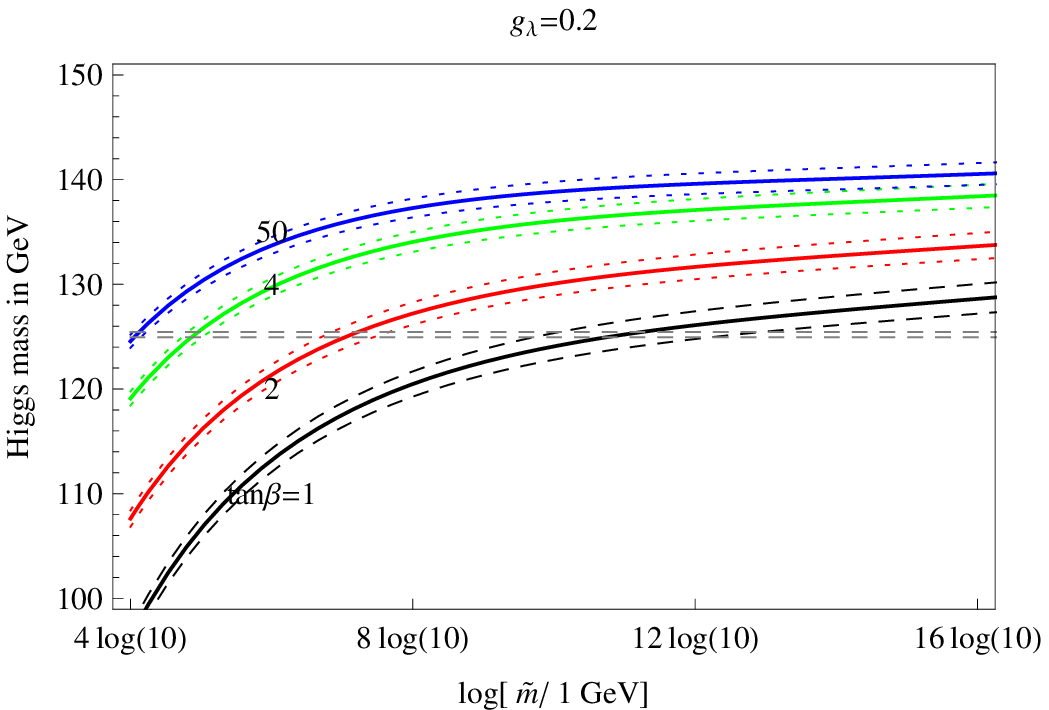}
\end{minipage}%
\centering
\begin{minipage}[b]{0.5\textwidth}
\centering
\includegraphics[width=3in]{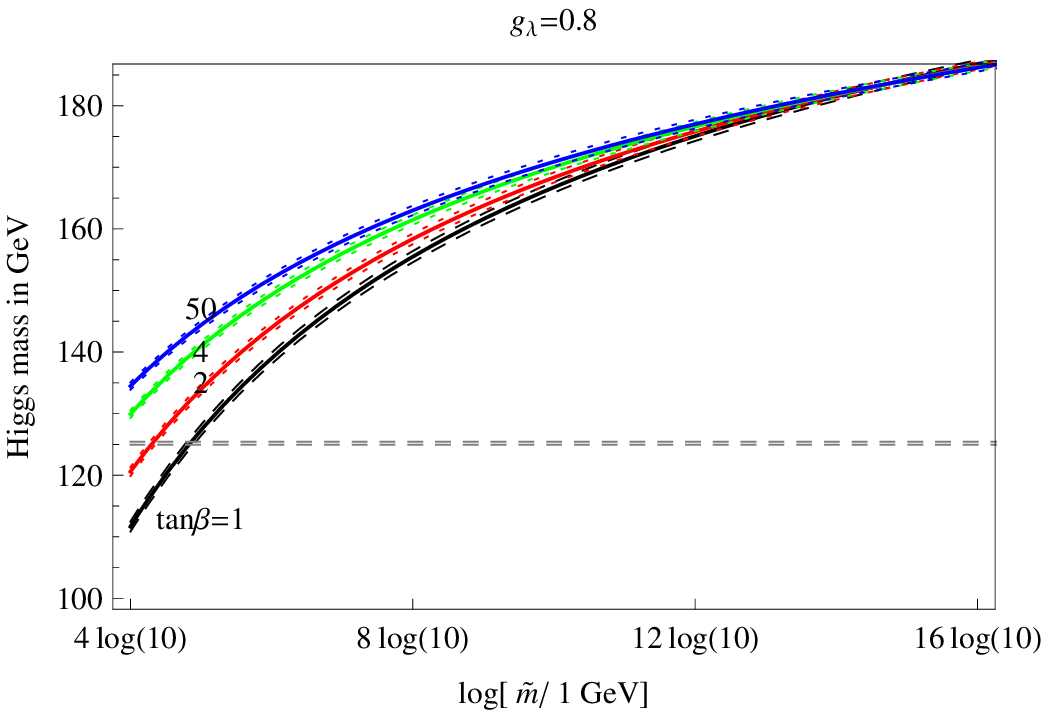}
\end{minipage}%
\caption{Higgs mass as function of SUSY-breaking scale in SM$+\tilde{h}/\tilde{s}$ by using two-loop RGEs, 
for  $\tan\beta=\{1, 2, 4, 50\}$ and $g_{\lambda}(M_{Z})=0.2$ (left) and $g_{\lambda}(M_{Z})=0.8$ (right).
Here each solid curve corresponds to the central values $M_{t}=173.3$ GeV and $\alpha_{3}(M_{Z})=0.1184$,
while the dotted curves in each color show uncertainty of Higgs mass due to experimental uncertainty in $m_{t}$.
Threshold corrections are ignored. 
}
\label{higgsino1}
\end{figure}

In model SM$+\tilde{h}/\tilde{s}$ 
two new parameters enter in the effective Lagrangian below $\tilde{m}$ \cite{0912.3942},
\begin{eqnarray}{\label{lagrangian}}
\mathcal{L}=\mathcal{L}_{SM}(q,u,d,l,e,h)+\mu\tilde{h}_{u}\tilde{h}_{d}
+\frac{m}{2}\tilde{s}^{2}+g_{\lambda}\tilde{h}_{d}\tilde{s} h+\text{hc}.
\end{eqnarray}
Ref. \cite{0912.3942} has shown that the observed DM relic abundance can be explained in the wide range $0<g_{\lambda}<0.9$.
As $\lambda(\tilde{m})$ is sensitive to  $g_{\lambda}(M_{Z})$,
we choose $g_{\lambda}=0.2$ in the small $g_{\lambda}$ region ($\leq0.4$)
and  $g_{\lambda}=0.8$ in the large $g_{\lambda}$ region ($\geq 0.7$) for comparison.

Fig. \ref{higgsino1} shows that the uncertainty of prediction for the Higgs mass at high energy scale is suppressed at EW scale, and there is only about $\sim 1-2$ GeV uncertainty  due to uncertainty of $m_{t}$.
For $g_{\lambda}=0.2$ (left panel)  it shows that the model is excluded for $\tilde{m}\geq 10^{10}$ GeV, 
while the model is excluded for $\tilde{m}\geq 10^{5}$ GeV instead for $g_{\lambda}=0.8$ (right panel). 
This obviously differs from the case for SM $+\tilde{w}$.
It is because that the deviation from SM in this model is larger than in SM $+\tilde{w}$,
especially in the large $g_{\lambda}$ region.
This attributes to the fact that $g_{\lambda}$-induced contribution to one-loop beta function $\Delta\beta^{(1)}_{\lambda}$ is negative in the large $g_{\lambda}$ region.

\linespread{1}\begin{figure}[!h]
\centering
\begin{minipage}[b]{0.5\textwidth}
\centering
\includegraphics[width=3in]{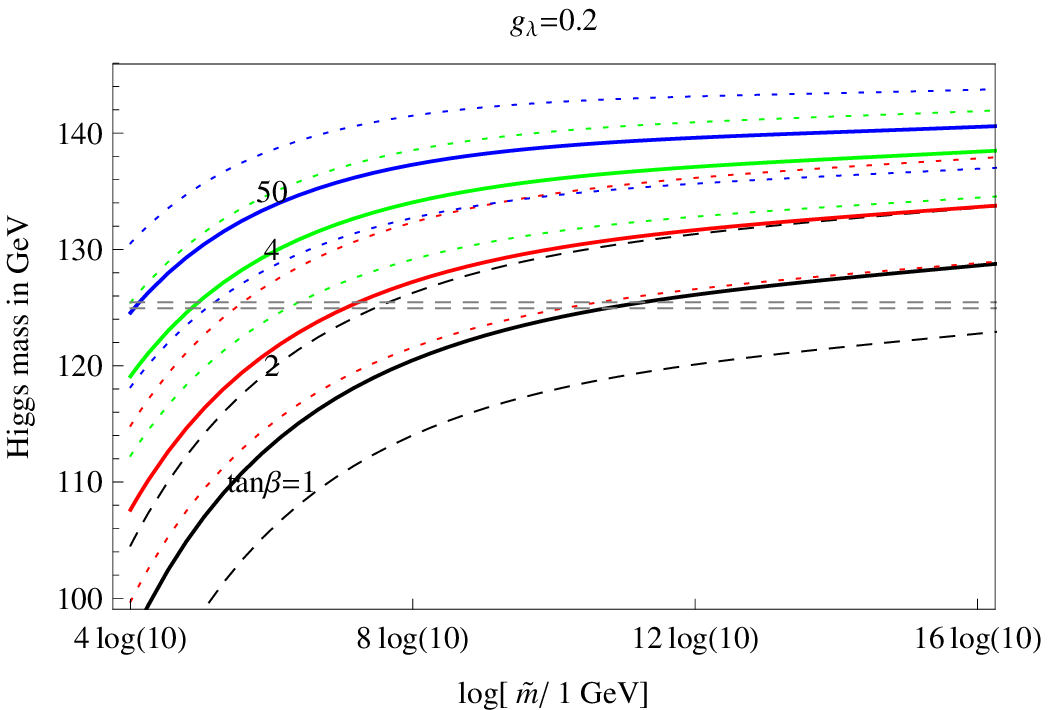}
\end{minipage}%
\centering
\begin{minipage}[b]{0.5\textwidth}
\centering
\includegraphics[width=3in]{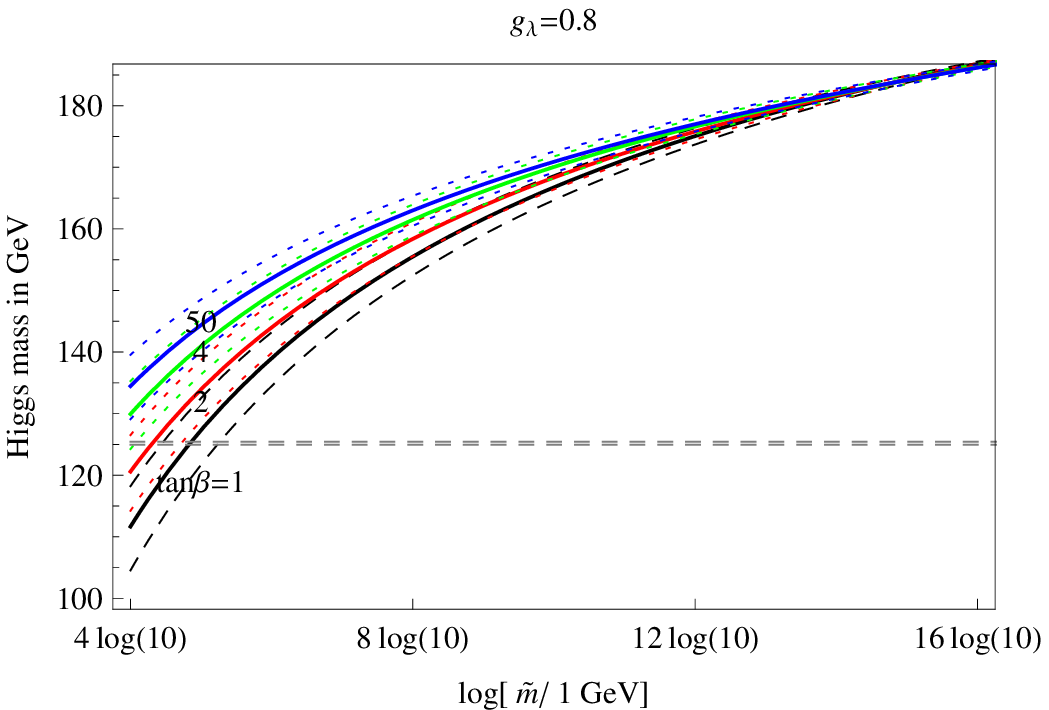}
\end{minipage}%
\caption{Same as Fig.\ref{higgsino1} but with threshold corrections $\mid\delta\lambda^{(\text{SUSY})}\mid \leq 0.03$.
The solid line in each color corresponds to the central values of $M_{t}(M_{Z})$ and $\alpha_{3}(M_{Z})$ and $\delta\lambda^{(\text{SUSY})}=0$.
Dotted lines in each color represent deviation from above due to the uncertainty of $M_{t}(M_{Z})$ and $\delta\lambda^{(\text{SUSY})}$.
}
\label{higgsino2}
\end{figure}

Fig.\ref{higgsino2} shows the case for threshold correction taken into account.
The solid line in each color corresponds to the central values of $M_{t}(M_{Z})$ and $\alpha_{3}(M_{Z})$ and $\delta\lambda^{(\text{SUSY})}=0$.
The dotted lines in each color represent the deviation from above due to the uncertainty of $M_{t}(M_{Z})$ and $\delta\lambda^{(\text{SUSY})}\neq 0$.
In comparison with Fig. \ref{higgsino1}  
it shows that in the small $g_{\lambda}$ region, 
the model is allowed for $\tilde{m}\leq 10^{7}$ GeV and $\tilde{m}\leq 10^{16}$ GeV
for threshold corrections $\delta\lambda^{(\text{SUSY})} = 0.03$ and $\delta\lambda^{(\text{SUSY})}=-0.03$, respectively.
However, in the large $g_{\lambda}$ region, 
the model is only allowed for $\tilde{m}\leq 10^{4}$ GeV and $\tilde{m}\leq 10^{6}$ GeV
for threshold corrections $\delta\lambda^{(\text{SUSY})}= 0.03$ and $\delta\lambda^{(\text{SUSY})} =-0.03$, respectively.
The combination of Fig.\ref{higgsino1} and Fig.\ref{higgsino2} implies that 
high-scale SUSY is not favored in the large  $g_{\lambda}$  region 
due to the significant and positive contribution to the Higgs quartic coupling $\lambda(M_{Z})$.

\section{Model Building}
This section is devoted to the model building of high-scale SUSY studied in the previous section.
We will show that high-scale SUSY which is unnatural from the viewpoint of EW scale
may be constructed at high energy scale.

\subsection{SM$+\tilde{w}$}
Firstly we employ gauge mediation  \footnote{For modern review on this topic, see, e.g., \cite{review}. 
Despite GM there is another important mediation mechanism of SUSY breaking referred as gravity mediation.
In gravity mediation, the wino mass is of the same order as the other two gaugino masses,
thus this framework can not be directly applied to our purpose. } (GM)
to construct effective theory below scale $\tilde{m}$ represented by SM$+\tilde{w}$.
Consider that the gluino and bino masses are far bigger than the wino mass in this model,
the messenger sector must be rather different from the setting as conventional GM,
in the later of which all of gaugino masses are of the same order.
Based on the messenger sector of conventional GM,
we consider a type of models as shown in table 1.
Obviously, the messengers in the model complete the $SU(5)$ representation of SM gauge symmetry.

\begin{table}
\begin{center}
\begin{tabular}{|c|c|}
  \hline
   & $SU(3)_{c}\times SU(2)_{L}\times U(1)_{Y}$ \\
  \hline
  $q$ & $\left(\mathbf{3}, 1, -\frac{1}{3}\right)$ \\
  $\bar{q}$ & $\left(\bar{\mathbf{3}}, 1, +\frac{1}{3}\right)$  \\
  $l$ & $\left(\mathbf{1}, \mathbf{2}, +\frac{1}{2}\right)$   \\
  $\bar{l}$ & $\left(\mathbf{1}, \bar{\mathbf{2}}, -\frac{1}{2}\right)$ \\
  \hline
\end{tabular}
\caption{The representations of messengers in the scenario under the SM gauge symmetry. }
\end{center}
\end{table}

We assume that the colored messengers ($q$, $\bar{q}$) and un-colored messengers ($l$, $\bar{l}$) in the model
couple to two different SUSY-breaking sector $X_{q}$ and $X_l$, respectively,
\begin{eqnarray}{\label{s1}}
W=X_{q} q\bar{q}+X_{l} l\bar{l},
\end{eqnarray}
where 
\begin{eqnarray}{\label{X}}
X_{q}=M_{q}+\theta^{2}F_{q}, ~~~~~~~~
X_{l}=M_{l}+\theta^{2}F_{l}.
\end{eqnarray}
Here $F_{q,l}$ and $M_{q,l}$ refer to the SUSY-breaking and tree-level mass scale for messengers involved, 
respectively.
For our purpose, we take the assumption that $M_{q}<<M_{l}$.

The scale arrangement in Eq.(\ref{X}) indicates that 
the mass scale $\tilde{m}$ and $m_{\tilde{w}}$ ($\sim$ EW scale)
can be identified as two different $F_{q}/M_{q}$ and $F_{l}/M_{l}$, respectively,
which is obvious in terms of the soft mass spectrum,
\begin{eqnarray}{\label{scalarmass}}
m^{2}_{\tilde{Q}}[M_{q}]&\simeq& \mathcal{A}\times \left(\frac{4}{3}\alpha^{2}_{3}[M_{q}]+
\frac{1}{60}\alpha^{2}_{1}[M_{q}]\right)+\mathcal{O}\left(\frac{F^{2}_{l}}{M^{2}_{l}}\right),\nonumber\\
 m^{2}_{\tilde{U}}[M_{q}]&\simeq&\mathcal{A}\times \left(\frac{4}{3}\alpha^{2}_{3}[M_{q}]+
\frac{4}{15}\alpha^{2}_{1}[M_{q}]\right)+\mathcal{O}\left(\frac{F^{2}_{l}}{M^{2}_{l}}\right),\nonumber\\
m^{2}_{\tilde{D}}[M_{q}]&\simeq& \mathcal{A}\times \left(\frac{4}{3}\alpha^{2}_{3}[M_{q}]+
\frac{1}{15}\alpha^{2}_{1}[M_{q}]\right)+\mathcal{O}\left(\frac{F^{2}_{l}}{M^{2}_{l}}\right),\nonumber\\
m^{2}_{\tilde{L}}[M_{q}]&\simeq& \mathcal{A}\times \left(\frac{3}{20}\alpha^{2}_{1}[M_{q}]\right)+\mathcal{O}\left(\frac{F^{2}_{l}}{M^{2}_{l}}\right),\nonumber\\
 m^{2}_{\tilde{E}}[M_{q}]&\simeq&\mathcal{A}\times \left(\frac{3}{5}\alpha^{2}_{1}[M_{q}]\right)+
\mathcal{O}\left(\frac{F^{2}_{l}}{M^{2}_{l}}\right),\nonumber\\
m^{2}_{H_{u}}[M_{q}]&\simeq& \mathcal{A}\times \left(\frac{3}{10}\alpha^{2}_{1}[M_{q}]\right)+\mathcal{O}\left(\frac{F^{2}_{l}}{M^{2}_{l}}\right),\nonumber\\
m^{2}_{H_{d}}[M_{q}]&\simeq& \mathcal{A}\times \left(\frac{3}{10}\alpha^{2}_{1}[M_{q}]\right)+\mathcal{O}\left(\frac{F^{2}_{l}}{M^{2}_{l}}\right).
\end{eqnarray}
Here $\mathcal{A}=\frac{1}{8\pi^{2}}\frac{F_{q}^{2}}{M_{q}^{2}}$, 
which determines the overall magnitude of above soft masses.
Similar to the soft scalar masses the gluino and bino mass 
appear at one-loop level of order $\mathcal{O}(F_{q}/M_{q})$,
while wino mass at one-loop level of order $\mathcal{O}(F_{l}/M_{l})$,
\begin{eqnarray}{\label{gauginomass}}
m_{\tilde{g}}[M_{q}]&\simeq& 3\cdot \frac{\alpha_{3}[M_{q}]}{4\pi} \cdot\frac{F_{q}}{M_{q}},\nonumber\\
m_{\tilde{w}}[M_{l}]&\simeq& 2\cdot \frac{\alpha_{2}[M_{l}]}{4\pi} \cdot\frac{F_{l}}{M_{l}},\nonumber\\
m_{\tilde{b}}[M_{q}]&\simeq& 1\cdot \frac{\alpha_{1}[M_{q}]}{4\pi} \cdot\frac{F_{q}}{M_{q}}.
\end{eqnarray}
In terms of Eq.(\ref{scalarmass}) and Eq.(\ref{gauginomass}) 
one finds that 
$\tilde{m}$ and $m_{\tilde{w}}$ are related to
\begin{eqnarray}{\label{ms}}
\tilde{m}\rightarrow \frac{F_{q}}{M_{q}} ,~~~\text{and}~~~
m_{\tilde{w}} \rightarrow \frac{F_{l}}{M_{l}},
\end{eqnarray}
respectively.
Therefore, the input wino mass $\tilde{m}$ is far beneath the scale $\sim \tilde{m}$ 
if there is a significant mass hierarchy $F_{q}/M_{q} >> F_{l}/M_{l}$.
This hierarchy is insured when $M_{l}>>M_{q}$ and /or $F_{q}>>F_{l}$.

However, the mass splitting between  the SUSY mass spectrum of order $\tilde{m}$ and the weak scale wino DM mass $m_{\tilde{w}}[v]$
is constrained by possibly large quantum correction\footnote {We thank the referee for pointing out this to us.}.
Because the  correction to wino mass arising from SUSY particles,
which is  of two-loop order,
may be larger than the input  mass $m_{\tilde{w}}[M_{l}]$ of order $v$
when $\tilde{m}$ is far above the weak scale.
The situation critically depends on whether this two-loop correction is logarithmic function of $\tilde{m}$ \footnote{
Unlike the low-scale SUSY,  
in the literature there is less interest to the study of wino DM mass in unnatural high-scale SUSY.
This subject will be studied in detail elsewhere \cite{new}.}.
If so, the fine tuning is mild similar to the case of two-loop SUSY QCD correction to the gluino mass \cite{gluino}. 
Otherwise, a wino mass of weak scale can not be obtained for arbitrary $\tilde{m}$.
For example, for the case of linear dependence, $\delta m_{\tilde{w}}\sim \tilde{m}/(16\pi^{2})^{2}$,
 the effective theory described by SM $+\tilde{w}$ is viable only for $\tilde{m}$ beneath $\sim (16\pi^{2})^{2} m_{\tilde{w}}[v]\sim 10^{7}$ GeV.
It means that some part of the parameter space in Fig.\ref{wino1} and Fig.\ref{wino2} can be achieved.

Actually, the constraint on the magnitude of mass splitting stands
as long as the effective theory below scale $\tilde{m}$ is similar to the model of SM $+\tilde{w}$,
regardless of the mediation mechanism of SUSY breaking. 
In particular, the constraint in the case of non-logarithmic dependence on $\tilde{m}$ may be relaxed only in some subtle situation.
For example, when a dramatical cancellation occurs among contributions with different signs to the wino mass,
 a wino mass of order $\sim 2$ TeV may be still obtained, 
in which one pays the price of a large fine tuning.
To be honest, this may be the only choice for the effective theory described by  SM $+\tilde{w}$ 
when the SUSY correction to the wino mass is not logarithmic function of $\tilde{m}$.

Alternatively,  the soft mass spectrum above can be realized in terms of a single spurion field $X$ 
other than two as discussed above.
Assume that the mass matrix which appears in the messenger superpotential corresponds to the form,
\begin{eqnarray}{\label{matrix}}
\mathcal{M}=\left(
\begin{array}{ll}
X & M \\
M& 0 
\end{array}\right).
\end{eqnarray}
It turns out that gaugino mass $\tilde{m}\sim F_{X}\partial (\log\det{\mathcal{M}})/\partial X \sim 0$ at the one-loop level of order $\mathcal{O}(F/M)$ by following the fact $\det\mathcal{M}=\text{const}$ \cite{9705228}.
The leading contribution to wino mass may be at one-loop level of order $\mathcal(O)(F^{3}/M^{5})$.
It is expected that $\tilde{m}$ and $m_{\tilde{w}}$ are related to different orders of $F/M$,
\begin{eqnarray}{\label{ms}}
\tilde{m}\rightarrow \frac{F}{M} ,~~~\text{and}~~~
m_{\tilde{w}}\rightarrow \frac{F^{3}}{M^{5}},
\end{eqnarray}
from which $\tilde{m}$ can be far above the EW wino mass due to a small ratio $F/M^{2}$.
Similar to what happens in the two spurion fields,
effective theory described by SM$+\tilde{w}$ is viable 
if either the SUSY correction to wino mass is only logarithmic function of $\tilde{m}$, or
there exists a large cancellation among these contributions.

\subsection{SM$+\tilde{h}/\tilde{s}$}
Now we proceed to discuss the realization of SM$+\tilde{h}/\tilde{s}$ in high-scale SUSY by using gauge mediation for illustration.
An obvious way to produce weak-scale effective Lagrangian in Eq.(\ref{lagrangian}) is through 
introducing renormalizable superpotential, 
\begin{eqnarray}{\label{s}}
W=g SH_{u}H_{d}+\frac{\kappa}{3}S^{3},
\end{eqnarray}
where $H_{u,d}$ refer to Higgs doublet superfields and $S$ a SM singlet superfield.
$S$ is taken to directly couple to two messengers in the messenger superpotential,
\begin{eqnarray}{\label{m2}}
W=X(\bar{\phi}_{1}\phi_{1}+\bar{\phi}_{2}\phi_{2}+\dots)+\lambda S \phi_{2} \bar{\phi}_{1}.
\end{eqnarray}
As firstly noted in \cite{9706540}, 
we also impose a $Z_3$ symmetry under which all chiral superfields have charge $1/3$, but for $\phi_{1}$ and $\bar{\phi}_{2}$ which have charges $-1/3$ and for $X$ which is neutral.
This parity ensures that $X=M+F\theta^{2}$ doesn't mix with singlet $S$.

There are a few comments in order, regarding soft masses induced by Eq.(\ref{m2}).
First, Messengers ignored in Eq.(\ref{m2}) are simply assumed to dominate 
the contribution to soft scalar scalar and gaugino masses of order $\mathcal{O}(F/M)$, 
similar to the minimal GM.
As we will explain below, the correction to soft masses due to Yukawa coupling $\lambda$ is small.
So, $\tilde{m}$ is related to $F/M$.

Second, $S$ scalar and fermion masses $m_{s}$ and $m_{\tilde{s}}$ due to Yukawa coupling $\lambda$ in Eq.(\ref{m2})
are given by, respectively,
\begin{eqnarray}{\label{ms}}
m^{2}_{s}&\sim&\frac{\alpha_{\lambda}}{8\pi^{2}}\left(\frac{D}{2}\alpha_{\lambda}-C\alpha_{\text{SM}}\right)\frac{F^{2}}{M^{2}} \nonumber\\
m_{\tilde{s}}&\sim&\frac{\alpha_{\lambda}}{4\pi}\frac{F}{M}\sim \left(\frac{\alpha_{\lambda}}{\alpha_{\text{SM}}}\right)\tilde{m}.
\end{eqnarray}
where $C$ and $D$ are two positive real numbers of order one \cite{9706540}.
We have used $\alpha_{\text{SM}}$ to represent the structure constant of SM gauge coupling.
If we don't introduce direct Yukawa coupling $\lambda$,
$m_{\tilde{s}}$ would vanish for $\lambda \rightarrow 0$.
The requirement $m_{\tilde{s}}\sim \mathcal{O}(v)$ leads to
\begin{eqnarray}{\label{smass}}
\alpha_{\lambda}&\sim& \left(\frac{v}{\tilde{m}}\right)\alpha_{\text{SM}},\nonumber\\
m^{2}_{s}&\sim&-\left(\frac{\alpha_{\lambda}}{\alpha_{\text{SM}}}\right)\tilde{m}^{2}\sim- v\cdot\tilde{m}.
\end{eqnarray}

Finally, it is crucial to note that the sign of $m^{2}_{s}$ is negative. 
Otherwise, it is impossible to develop a non-zero vacuum expectation value for $S$ scalar,
and effective $\mu=\lambda\left< s\right>$ vanishes.
Recall that  the scalar potential for $s$ in the region of large $\left< s\right>$ is given by \cite{0910.1785},
\begin{eqnarray}{\label{po}}
V(s)\simeq m^{2}_{s}s^{2}+\frac{2}{3}\kappa A_{\kappa} s^{3}+\kappa^{2} s^{4}.
\end{eqnarray}
With negative $m^{2}_{s}$ in Eq.(\ref{smass}) and one-loop order $A_{\kappa}$ term, 
with $A_{\kappa}\sim -m_{\tilde{s}}$ \cite{1007.3323} 
we obtain the effective $\mu$ term,
\begin{eqnarray}{\label{mu}}
\mu\sim \frac{\lambda}{4\kappa} \left(-A_{\kappa}+\sqrt{A_{\kappa}^{2}-8m^{2}_{s}}\right)\sim \frac{\lambda}{\kappa}\sqrt{v\cdot \tilde{m}}.
\end{eqnarray}
So, $\mu$ term of weak scale implies that
\begin{eqnarray}
\frac{\lambda}{\kappa}\sim \sqrt{\frac{v}{\tilde{m}}} \sim \frac{\lambda}{g_{\text{SM}}}
\end{eqnarray}

In summary, high-scale SUSY with effective theory beneath $\tilde{m}$ described by SM$+\tilde{h}/\tilde{s}$
can be realized by adopting a small singlet-messenger-messenger Yukawa coupling of order $\left(\frac{v}{\tilde{m}}\right)^{1/2}g_{\text{SM}}$ and singlet self coupling of order $\sim g_{\text{SM}}$. 
The smallness of $\lambda$ can be understood as consequence of either environmental selection or symmetric principle.
The model building discussed so far is completed.
In the next section, 
we will discuss the possible constraint on high-scale SUSY from the viewpoint of inflation physics.

\section{Constraints in the Light of Inflation}
So far we have explored high-scale SUSY with new physics scale $\tilde{m}$ far above EW scale.
The collider phenomenology of such high-scale SUSY is very similar to a large class of models 
whose matter content below $\tilde{m}$ is composed of SM together with weak-scale DM.
In \cite{1203.5106} a detailed analysis has been dedicated to DM thermal relic abundance and direct detection in these models.
Here we propose another independent portal to explore high-scale SUSY.
Roughly speaking, the inflation physics in the early universe may be either related 
or decoupled to the SUSY breaking sector which is responsible to the high-scale SUSY of particle physics.
If it is indeed related \cite{assumption},
as we assume in this section, 
experiments such as WAMP, Plank and BICEP, 
which devote to measure Cosmic Microwave Background temperature anisotropy and polarization 
induced by inflation,
probably expose high-scale SUSY at energy scale near $\tilde{m}$ from the study of inflation physics
\footnote{For recent attempt to address high-scale SUSY in the light of physics of early universe, see, e.g., \cite{1409.7462}.}.

We restrict our study to single-field inflation, 
in which the scalar spectral index $n_{s}$ and tensor spectral index $n_{t}$ are given by, respectively,
\begin{eqnarray}{\label{index}}
n_{s}-1\simeq 2\eta-6\epsilon,~~~~~~~~~
n_{t}\simeq -2\epsilon.
\end{eqnarray}
Here slow roll parameter $\eta=M^{2}_{P}V_{,\phi\phi}/V$ and $\epsilon=\frac{M^{2}_{P}}{2}\left(V_{,\phi}/V\right)^{2}$, with subscript denoting derivative of inflation potential $V$ over
inflaton field $\phi$. 
$M_{P}=1.0\times 10^{19}$ GeV is the Plank mass.
Among other things, 
$n_{s}$ and the scalar-to-tensor ratio $r$ are the most important quantities \cite{1403.6310} to temporary study of inflation physics.
In terms of these two quantities, 
which are measured at WAMP, Plank and BICEP experiments, 
high-scale mass scale $\tilde{m}\sim m_{\phi}$ can be directly probed.

\linespread{1}\begin{figure}[!h]
\centering
\includegraphics[width=2.7in]{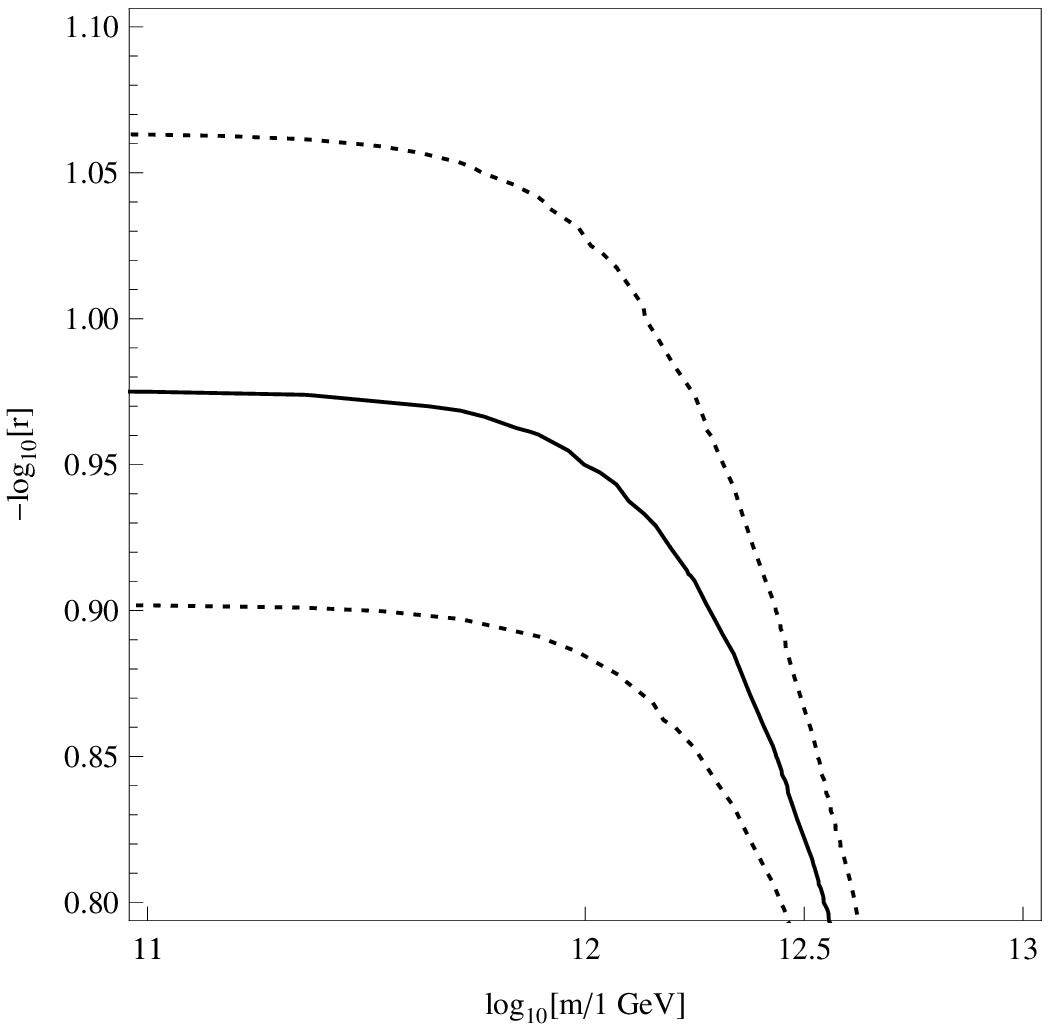}
\includegraphics[width=2.7in]{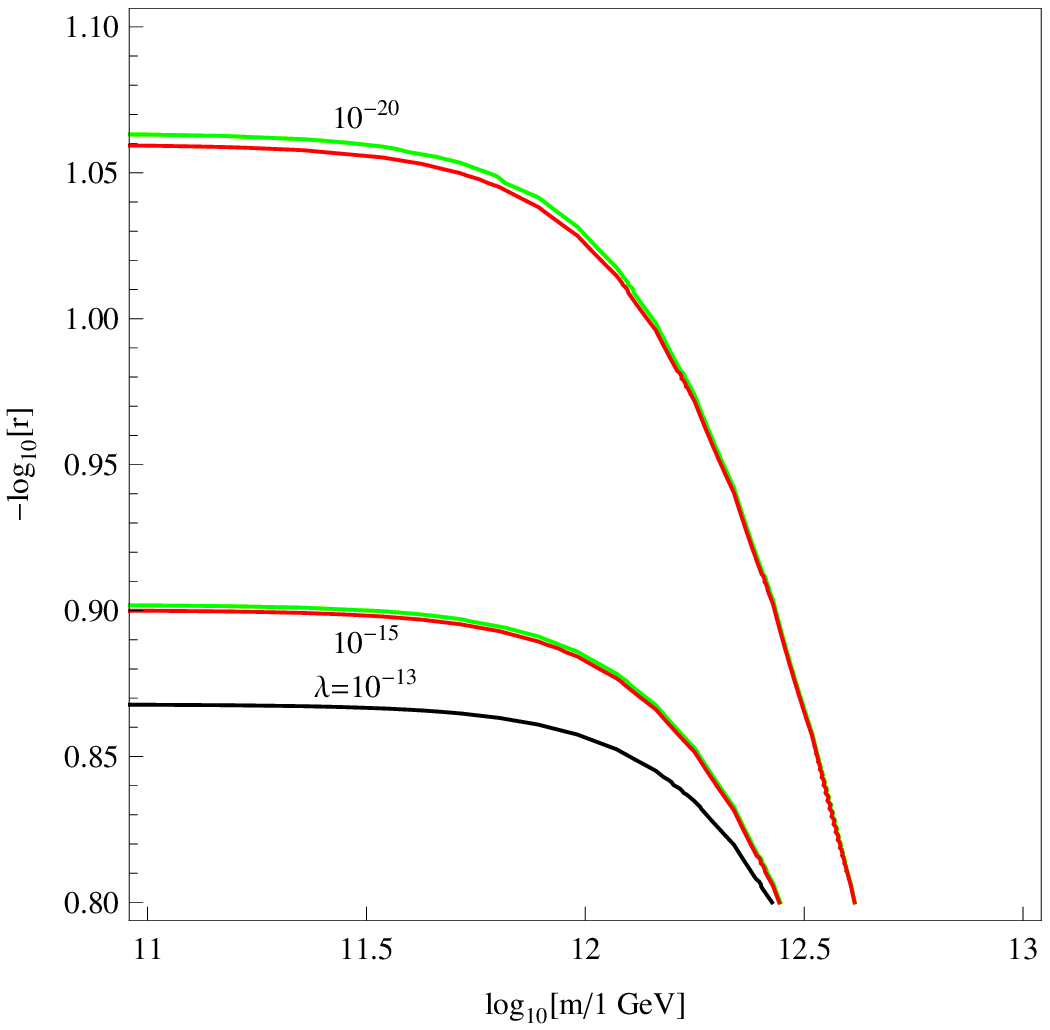}
\centering
\caption{Contour of $n_s$ in the plane of $(\tilde{m}, r)$ for quadratic (left) and quartic (right) inflation potential.
In the left panel, solid line corresponds to the central value of $n_s$ and dotted lines show the uncertainty due to the uncertainty of $n_{s}$. 
In the right panel, we show the contours for $\left<\phi\right>=M_{P}$ and $\lambda_{\phi}=\{10^{-13},10^{-15},10^{-20}\}$. 
Note that small slow parameters $\eta$ and $\epsilon$ require $\lambda<<5\cdot10^{-12}$.}
\label{in}
\end{figure}

The combination of nine-year WAMP and Plank data leads to
$n_{s}=0.9603\pm 0.0073$ at 68\% CL \cite{1303.5082}.
Unlike $n_s$, the measured value for $r$ isn't in high precision at present status.
The combination of WAMP and Plank data has reported that $r< 0.11$ at 95\% CL \cite{1303.5082}, while the BICEP2 reported a relatively large value $r\sim 0.16$ \cite{1403.3985}.
Consider chaotic inflation potential 
\begin{eqnarray}{\label{potential}}
V(\phi)=V_{0}+\frac{1}{2}m^{2}_{\phi}\phi^{2}+\frac{\lambda_{\phi}}{12}\phi^{4},
\end{eqnarray}
where $V_0$ is the vacuum energy due to SUSY breaking.
For the case of single-field inflation 
substituting $\epsilon=r/16$ and effective value $V^{1/4}\simeq 2\times 10^{16} \cdot \left(r/0.20\right)^{1/4}$ GeV  into Eq.\eqref{index} 
we find,
\begin{eqnarray}{\label{c}}
n_{s}\simeq 1+\left(\frac{r}{0.1}\right)^{-1}\left(\frac{f(\tilde{m})}{2\cdot 10^{13}~ \text{GeV}}\right)^{2}-0.0375\cdot\left(\frac{r}{0.1}\right).
\end{eqnarray}
with $f^{2}(\tilde{m})=\partial^{2} V/\partial \phi^{2}$, which reads as,
\begin{eqnarray}{\label{f}}
f^{2}(\tilde{m})=\begin{cases}
\tilde{m}^{2}& (\text{quadratic})\\
\tilde{m}^{2}+\lambda_{\phi}\left<\phi\right>^{2} & (\text{quartic})
\end{cases}
\end{eqnarray}
In terms of Eq.\eqref{f} one can derive the bound on $\tilde{m}$ by using measured values of $n_s$ and $r$.

In Fig.\ref{in} we show the bound on $\tilde{m}$ for quadratic (left) and quartic (right) inflation potential.
The left panel solid line corresponds to the central value of $n_s$ 
and dotted lines show the uncertainty due to the uncertainty of $n_{s}$. 
It shows that $\tilde{m}>10^{13}$ GeV is excluded for $-\log_{10}(r)$ in the range of $[0.9, 1.06]$.
In the right panel, similar bound holds for $\left<\phi\right>=M_{P}$ and different values of $\lambda_{s}$.
The model is excluded for $\lambda\geq10^{-12}$ or $-\log_{10}(r)$ outside the range of $[0.87, 1.06]$.
Small $r<<0.1$ is excluded in these two simple inflation models due to the fact that observed $n_{s}$ is smaller than unity.

\section{Conclusions}
Inspired by the present LHC results on SUSY,
the prediction for the Higgs mass in high-scale SUSY with weak interacting massive DM is explored in this paper.
Similar to well known split SUSY, 
models of SM $+\tilde{w}$ and SM $+\tilde{h}/\tilde{s}$,
in which wino and mixing state of higgsino and singlino serves as DM respectively, 
are studied in detail.
The main results in this study include: 
$i)$, in model of SM $+\tilde{w}$, the SUSY-breaking scale $\tilde{m}$ is allowed in 
the whole range of $10^{4}$ GeV $\leq \tilde{m}\leq 10^{16}$ GeV for vanishing threshold correction,
and $\tilde{m} \leq 10^{8}$ GeV is still allowed for $ \delta \lambda_{max} = 0.03$.
$ii)$, in model of SM $+\tilde{h}/\tilde{s}$ with vanishing threshold correction $\delta\lambda^{(\text{SUSY})}=0$, 
$\tilde{m}\leq 10^{10}$ ($10^{5}$ ) GeV is allowed in the small (large) $g_{\lambda}$ region. 
For threshold corrections $\delta\lambda^{(\text{SUSY})}_{max} = 0.03$,
the model is still allowed for $\tilde{m}\leq 10^{7}$ ($10^{4}$) GeV in the small (large) $g_{\lambda}$ region.

As for the model building of high-scale SUSY studied here,
we propose that  the effective theory below scale $\tilde{m}$ described by SM $+\tilde{w}$,  
can be realized in the non-minimal GM with two spurion fields that exhibit significant mass hierarchy.
We also mention that the bound on $\tilde{m}$ may dramatically decrease 
when the two-loop SUSY correction to the wino DM  mass is not logarithmic function of $\tilde{m}$.
On the other hand, the effective theory below scale $\tilde{m}$ described by SM $+\tilde{h}/\tilde{s}$,
can be realized in GM with direct singlet-messenger-messenger coupling for
the magnitude of Yukawa coupling relative to SM gauge coupling $g_{\text{SM}}$ of order $\sim(v/\tilde{m})^{1/2}$.

Since high-scale SUSY loses its connection to the weak scale, 
and also has no promising prospect for discovery at the LHC,  
we argue that the prediction for the Higgs mass encourages relating it to cosmology of the early universe, 
especially the inflation physics,
which constrains $\tilde{m}<10^{13}$ GeV in the quadratic or quartic inflation potential with $r\simeq 0.1$.

\begin{acknowledgments}
The author thanks Y. Bai and M. Luo for useful discussions,
and Y. Ma for the help of numerical simulation.
The work is supported in part by Natural Science Foundation of China under grant No.11247031 and 11405015.
\end{acknowledgments}

\newpage

\appendix
\section{RGE}
Given a coupling $g_i$, its RGE can be written as, 
\begin{eqnarray}{\label{def}}
\frac{d g_{i}}{d\ln Q}=\frac{\beta^{(1)}(g_{i})}{(4\pi)^{2}} +\frac{\beta^{(2)}(g_{i})}{(4\pi)^{4}}
\end{eqnarray}
where $\beta^{(1)}(g_{i})$ and $\beta^{(2)}(g_{i})$ is the one- and two-loop beta function for $g_i$, respectively.
$g_{i}$ denote all relevant couplings. 
The beta functions for SM are simply denoted by $\beta^{(1)}$ and $\beta^{(2)}$,
with the corrections to them due to new physics referred by $\Delta\beta^{(1)}$ and $\Delta\beta^{(2)}$, respectively.

$\mathbf{SM}$\\
With SM Yukawa couplings except top Yukawa $y_t$ ignored the RGEs for SM read as,
\begin{eqnarray}{\label{sm}}
\beta^{(1)}(g_{1})&=&\frac{41}{10}g^{3}_{1} \nonumber\\
\beta^{(2)}(g_{1})&=&-\frac{17}{10}g^{3}_{1}y^{2}_{t} +\frac{199}{50}g^{5}_{1}+\frac{27}{10}g_{1}^{3}g_{2}^{2}+\frac{44}{5}g_{1}^{3}g_{3}^{2}\nonumber\\
\beta^{(1)}(g_{2})&=&-\frac{19}{6}g^{3}_{2} \nonumber\\
\beta^{(2)}(g_{2})&=&-\frac{3}{2}g^{3}_{2}y^{2}_{t} +\frac{35}{6}g^{5}_{2}+\frac{9}{10}g_{1}^{2}g_{2}^{3}+12 g_{2}^{3}g_{3}^{2}\nonumber\\
\beta^{(1)}(g_{3})&=&-7g^{3}_{3} \nonumber\\
\beta^{(2)}(g_{3})&=&-2g^{3}_{1}y^{2}_{t} -26g^{5}_{3}+\frac{11}{10}g_{1}^{2}g_{3}^{3}+\frac{9}{2}g_{3}^{3}g_{2}^{2} \nonumber\\
\beta^{(1)}(y_{t})&=&\frac{9}{2}y_{t}^{3}-\frac{17}{20}y_{t}g^{2}_{1}-\frac{9}{4}y_{t}g^{2}_{2}-8y_{t}g^{2}_{3}\nonumber\\
\beta^{(2)}(y_{t})&=&\frac{393}{80}g^{2}_{1}y^{3}_{t}+\frac{225}{16}g^{2}_{2}y^{3}_{t}+36g^{2}_{3}y^{3}_{t}+\frac{1187}{600}g^{4}_{1}y_{t}-\frac{23}{4}g^{4}_{2}y_{t}-108g^{4}_{3}y_{t}-\frac{9}{20}g^{2}_{1}g^{2}_{2}y_{t}\nonumber\\
&+&\frac{19}{15}g^{2}_{1}g^{2}_{3}y_{t}+9g^{2}_{2}g^{2}_{3}y_{t}-6\lambda~y^{3}_{t}+\frac{3}{2}\lambda^{2}y_{t}-12y^{5}_{t}
\end{eqnarray}
 $\mathbf{SM+ \tilde{w}}$\\
The corrections $\Delta\beta^{(1)}$ and $\Delta\beta^{(2)}$ in $\text{SM}+\tilde{w}$ can be derived by decoupling the Dirac doublet fermion in models studied in \cite{1203.5106}, which are given by,
\begin{eqnarray}{\label{smw}}
\Delta\beta^{(2)}(g_{1})&=&\frac{9}{50}g^{5}_{1}+\frac{9}{10}g_{1}^{3}g_{2}^{2}\nonumber\\
\Delta\beta^{(1)}(g_{2})&=&\frac{4}{3}g^{3}_{2} \nonumber\\
\Delta\beta^{(2)}(g_{2})&=&\frac{59}{2}g^{5}_{2}+\frac{3}{10}g_{1}^{2}g_{2}^{3}\nonumber\\
\Delta\beta^{(2)}(y_{t})&=&\frac{29}{150}g^{4}_{1}y_{t}+\frac{3}{2}g^{4}_{2}y_{t}
\end{eqnarray}
$\mathbf{SM+ \tilde{h}/\tilde{s}}$\\
Two new Yukawa couplings $g_{\lambda}$ and $\tilde{g}_{\lambda}$ 
which appear in interaction terms $g_{\lambda}h\tilde{s}\tilde{h}_{d}$ and $\tilde{g}_{\lambda}h\tilde{s}\tilde{h}_{u}$ 
enter in SM$+\tilde{h}/\tilde{s}$.
Similarly to $\text{SM}+\tilde{w}$, the corrections to beta functions can be derived in terms of mathching our model to SM $+$ Majorana singlet fermion $+$ Dirac doublet fermion (D) in \cite{1203.5106}.
Identify $D$ and $D^{c}$ as higgsinos, we obtain in the small $\tilde{g}_{\lambda}$ region, 
\begin{eqnarray}{\label{smhs}}
\Delta\beta^{(1)}(g_{\lambda})&=&g_{\lambda}\left(\frac{5}{2}g^{2}_{\lambda}+3y^{2}_{t}-\frac{9}{4}g^{2}_{2}-\frac{9}{20}g^{2}_{1}\right)\nonumber\\
\Delta\beta^{(1)}(g_{1})&=&\frac{2}{5}g^{3}_{1}\nonumber\\
\Delta\beta^{(2)}(g_{1})&=&\frac{3}{10}g^{3}_{1}(g^{2}_{\lambda}+\tilde{g}^{2}_{\lambda})+\frac{9}{50}g^{5}_{1}+\frac{9}{10}g^{2}_{2}g^{3}_{1}\nonumber\\
\Delta\beta^{(2)}(g_{2})&=&-\frac{1}{2}g^{3}_{2}(g^{2}_{\lambda}+\tilde{g}^{2}_{\lambda})+\frac{49}{6}g^{5}_{2}+\frac{3}{10}g^{2}_{1}g^{3}_{2}\\
\Delta\beta^{(1)}(y_{t})&=&y_{t}(g^{2}_{\lambda}+\tilde{g}^{2}_{\lambda})\nonumber\\
\Delta\beta^{(2)}(y_{t})&=&(g^{2}_{\lambda}+\tilde{g}^{2}_{\lambda})(\frac{3}{8}g^{2}_{1}y_{t}+\frac{15}{8}g^{2}_{2}y_{t}-\frac{9}{4}y^{3}_{t})+\frac{29}{150}g^{4}_{1}y_{t}+\frac{1}{2}g^{4}_{2}y_{t}-\frac{9}{4}y_{t}(g^{4}_{\lambda}+\tilde{g}^{4}_{\lambda})-5y_{t}g^{2}_{\lambda}\tilde{g}^{2}_{\lambda}\nonumber
\end{eqnarray}
$\mathbf{Higgs~quartic}$\\
Finally, the two-loop RGE for Higgs quartic $\lambda$ in SM$+\tilde{w}$ is given by,
\begin{eqnarray}{\label{quartic1}}
\beta^{(1)}(\lambda)&=&\lambda(12\lambda+12y^{2}_{t}-\frac{9}{5}g^{2}_{1}-9g^{2}_{2})
-12y^{4}_{t}+\frac{27}{100} g^{4}_{1}+\frac{9}{10}g^{2}_{1}g^{2}_{2}+\frac{9}{4}g^{4}_{2}\nonumber\\
\beta^{(2)}(\lambda)&=&-78\lambda^{3}+\lambda^{2}\left(\frac{54}{5}g^{2}_{1}+54g^{2}_{2}-72y^{2}_{t}\right)\nonumber\\
&+&
\lambda\left[\frac{1887}{200}g^{4}_{1}+\frac{117}{20}g^{2}_{1}g^{2}_{2}-\frac{73}{8}g^{4}_{2}-3 y^{4}_{t}+g^{2}_{t}\left(\frac{17}{2}g^{2}_{1}+\frac{45}{2}g^{2}_{2}+80g^{2}_{3}\right)\right]
\nonumber\\
&+&60y^{6}_{t}-\frac{3411}{1000}g^{6}_{1}+\frac{305}{8}g^{6}_{2}-\frac{289}{40}g^{2}_{1}g^{4}_{2}-\frac{1773}{200}g^{4}_{1}g_{2}^{2}-64g^{2}_{3}y^{4}_{t}-\frac{16}{5}g^{2}_{1}y^{4}_{t}-\frac{9}{2}g^{4}_{2}y^{2}_{t}\nonumber\\
&+&\frac{3}{5}g^{2}_{1}y^{2}_{t}\left(-\frac{57}{10}g^{2}_{1}+21g^{2}_{2}\right)
+g^{4}_{2}\left(-12g^{2}_{2}+15\lambda-\frac{12}{5}g^{2}_{1}\right).
\end{eqnarray}
while in SM$+\tilde{h}/\tilde{s}$ reads as,
\begin{eqnarray}{\label{quartic2}}
\beta^{(1)}(\lambda)&=&\lambda(12\lambda+12y^{2}_{t}-\frac{9}{5}g^{2}_{1}-9g^{2}_{2})
-12y^{4}_{t}+\frac{27}{100} g^{4}_{1}+\frac{9}{10}g^{2}_{1}g^{2}_{2}+\frac{9}{4}g^{4}_{2}
+4\lambda g^{2}_{\lambda}-4g_{\lambda}^{4}\nonumber\\
\beta^{(2)}(\lambda)&=&-78\lambda^{3}+\lambda^{2}\left(\frac{54}{5}g^{2}_{1}+54g^{2}_{2}-72y^{2}_{t}\right)+\frac{3}{5}g^{2}_{1}y^{2}_{t}\left(-\frac{57}{10}g^{2}_{1}+21g^{2}_{2}\right)\nonumber\\
&+&
\lambda\left[\frac{1887}{200}g^{4}_{1}+\frac{117}{20}g^{2}_{1}g^{2}_{2}-\frac{73}{8}g^{4}_{2}-3 y^{4}_{t}+g^{2}_{t}\left(\frac{17}{2}g^{2}_{1}+\frac{45}{2}g^{2}_{2}+80g^{2}_{3}\right)\right]
\nonumber\\
&+&60y^{6}_{t}-\frac{3411}{1000}g^{6}_{1}+\frac{305}{8}g^{6}_{2}-\frac{289}{40}g^{2}_{1}g^{4}_{2}-\frac{1773}{200}g^{4}_{1}g_{2}^{2}-64g^{2}_{3}y^{4}_{t}-\frac{16}{5}g^{2}_{1}y^{4}_{t}-\frac{9}{2}g^{4}_{2}y^{2}_{t}\nonumber\\
&+&20g^{6}_{\lambda}-g^{4}_{\lambda}+2g^{2}_{\lambda}\left(-\frac{9}{100}g^{4}_{1}-12\lambda^{2}-\frac{3}{4}g^{4}_{2}-\frac{3}{10}g^{2}_{1}g^{2}_{2}+\frac{3}{4}\lambda g^{2}_{1}+\frac{15}{4}\lambda g^{2}_{2}\right)
\end{eqnarray}

\linespread{1}


\begin{thebibliography}{99}
\bibitem{splitsusy}
N.~ Arkani-Hamed and S.~Dimopoulos, 
JHEP {\bf 0506},~073~ (2005),  [hep-th/0405159];\\
G.~F.~Giudice and A.~Romanino, 
Nucl.\ Phys.\ B {\bf 699}, 65~(2004),  [hep-ph/0406088];\\
N.~Arkani-Hamed, S.~Dimopoulos, G.~F.~Giudice and A.~Romanino, 
Nucl.\ Phys.\ B {\bf 709}, 3~(2005),  [hep-ph/0409232].


\bibitem{higgsmass}
CMS Collaboration, Phys.\ Rev.\ D {\bf 89}, 092007 (2014), arXiv:1312.5353;\\
CMS Collaboration,  
Eur.\ Phys.\ J.\ C {\bf 74}, 3076 (2014), arXiv:1407.0558 [hep-ex] ;\\
ATLAS Collaboration, Phys.\ Rev.\ D {\bf 90}, 052004 (2014), arXiv:1406.3827[hep-ex] .


\bibitem{1108.6077}
 G.~F.~Giudice and A.~Strumia,
  %``Probing High-Scale and Split Supersymmetry with Higgs Mass Measurements,''
Nucl.\ Phys.\ B {\bf 858}, 63 (2012),
arXiv:1108.6077 [hep-ph].


\bibitem{1407.4081}
E.~Bagnaschi, G.~F.~Giudice, P.~Slavich and A.~Strumia,
 %``Higgs Mass and Unnatural Supersymmetry,''
arXiv:1407.4081 [hep-ph].


\bibitem{0912.3942}
G.~Elor, H.~-S.~Goh, L.~J.~Hall, P.~Kumar and Y.~Nomura,
 %``Environmentally Selected WIMP Dark Matter with High-Scale Supersymmetry Breaking,''
Phys.\ Rev.\ D {\bf 81}, 095003 (2010),
arXiv:0912.3942 [hep-ph].

\bibitem{0910.2235}
L.~J.~Hall and Y.~Nomura,
  %``A Finely-Predicted Higgs Boson Mass from A Finely-Tuned Weak Scale,''
JHEP {\bf 1003}, 076 (2010),
arXiv:0910.2235 [hep-ph].

\bibitem{threshold}
 A.~Sirlin and R.~Zucchini,
  %``Dependence of the Quartic Coupling H(m) on M($H$) and the Possible Onset of New Physics in the Higgs Sector of the Standard Model,''
Nucl.\ Phys.\ B {\bf 266}, 389 (1986).

\bibitem{1307.3536}
D.~Buttazzo, G.~Degrassi, P.~P.~Giardino, G.~F.~Giudice, F.~Sala, A.~Salvio and A.~Strumia,
%``Investigating the near-criticality of the Higgs boson,''
JHEP {\bf 1312}, 089 (2013),
arXiv:1307.3536 [hep-ph].

%%%%%%%%%%%%%%%% RGE for MSSM%%%%%%%%%%%%

\bibitem{Machacek}
M.~E.~Machacek and M.~T.~Vaughn,
Nucl.\ Phys.\ B {\bf236},~221~(1984); \\
M.~E.~Machacek and M.~T.~Vaughn, 
Nucl.\ Phys.\ B {\bf 249},~70~(1985);\\
M.~E.~Machacek and M.~T.~Vaughn,
Phys.\ Rev.\ D {\bf 50}, ~2282~(1994), [hep-ph/9311340].

\bibitem{Luo}
M.~Luo, Y.~Xiao, 
Phys.\ Rev.\ Lett.  {\bf 90},~011601~(2003),  [hep-ph/0207271]; \\
M.~Luo, H.~Wang, Y.~Xiao, 
Phys.\ Rev.\ D {\bf 67},~065019~(2003), [hep-ph/0211440].



\bibitem{weinberg}
Chapter 28 in S.~Weinberg,
 ``The quantum theory of fields. Vol. 3: Supersymmetry,''
Cambridge, UK: Univ. Pr. (2000) 419 p,  




%%%%%%%%%%%%%%%%%
\bibitem{1301.5167}
 L.~E.~Ibanez and I.~Valenzuela,
  %``The Higgs Mass as a Signature of Heavy SUSY,''
JHEP {\bf 1305}, 064 (2013),
arXiv:1301.5167 [hep-ph].

\bibitem{topmass}
The ATLAS, CDF, CMS and D0 Collaborations,  arXiv:1403.4427.


\bibitem{g3}
S.~Bethke, Eur.\ Phys.\ J.\ C {\bf 64},~689~(2009), 
arXiv:0908.1135.




\bibitem{review}
G.~F.~Giudice and  R.~Rattazzi,
Phys. Repts. {\bf322}~(1999)~419.


\bibitem{new}
S. Zheng et al,  in preparation.



\bibitem{gluino}
 Y.~Yamada,
  %``Two-loop SUSY QCD correction to the gluino pole mass,''
Phys.\ Lett.\ B {\bf 623}, 104 (2005),
[hep-ph/0506262].



\bibitem{9705228}
K.~I.~Izawa, Y.~Nomura, K.~Tobe and T.~Yanagida,
  %``Direct transmission models of dynamical supersymmetry breaking,''
Phys.\ Rev.\ D {\bf 56}, 2886 (1997),
[hep-ph/9705228].



\bibitem{9706540}
 G.~F.~Giudice and R.~Rattazzi,
  %``Extracting supersymmetry breaking effects from wave function renormalization,''
Nucl.\ Phys.\ B {\bf 511}, 25 (1998), 
[hep-ph/9706540].


\bibitem{0910.1785}
U.~Ellwanger, C.~Hugonie and A.~M.~Teixeira,
 ``The Next-to-Minimal Supersymmetric Standard Model,''
Phys.\ Rept.\  {\bf 496}, 1 (2010),
arXiv:0910.1785 [hep-ph].




\bibitem{1007.3323}
K.~Hamaguchi and N.~Yokozaki,
 %``Soft Leptogenesis and Gravitino Dark Matter in Gauge Mediation,''
Phys.\ Lett.\ B {\bf 694}, 398 (2011),
arXiv:1007.3323 [hep-ph].


\bibitem{1203.5106}
C.~Cheung, M.~Papucci and K.~M.~Zurek,
 %``Higgs and Dark Matter Hints of an Oasis in the Desert,''
JHEP {\bf 1207}, 105 (2012),
arXiv:1203.5106 [hep-ph].


\bibitem{1409.7462}
S.~Zheng,
 %``The Early Universe with High-Scale Supersymmetry,''
arXiv:1409.7462 [hep-ph].

\bibitem{1403.6310}
S.~Dodelson,
  %``How much can we learn about the physics of inflation?,''
Phys.\ Rev.\ Lett.\  {\bf 112}, 191301 (2014),
arXiv:1403.6310 [astro-ph.CO].


\bibitem{assumption}
 L.~J.~Hall, Y.~Nomura and S.~Shirai,
  %``Grand Unification, Axion, and Inflation in Intermediate Scale Supersymmetry,''
JHEP {\bf 1406}, 137 (2014),
arXiv:1403.8138 [hep-ph];\\
L.~E.~Ibanez and I.~Valenzuela,
 % ``BICEP2, the Higgs Mass and the SUSY-breaking Scale,''
arXiv:1403.6081 [hep-ph];\\
S.~Zheng,
 %``High-Scale SUSY Breaking as the Same Origin of Inflation After BICEP2 and Higgs Mass After LHC,''
arXiv:1405.2775 [hep-ph].



\bibitem{1303.5082}
Plank Collaboration,
%``Planck 2013 results. XXII. Constraints on inflation,''
arXiv:1303.5082 [astro-ph.CO].


\bibitem{1403.3985}
BICEP2 Collaboration,
% ``BICEP2 I: Detection Of B-mode Polarization at Degree Angular Scales,''
Phys.\ Rev.\ Lett.\  {\bf 112}, 241101 (2014),
arXiv:1403.3985 [astro-ph.CO].


\end{thebibliography}
\end{document}